\documentclass[journal=jacs,manuscript=communication]{achemso}
\usepackage[version=3]{mhchem} 
\usepackage{outlines}
\setkeys{acs}{layout = traditional}
\setkeys{acs}{usetitle = true}
\usepackage{graphicx,dcolumn,bm,amsmath,amssymb}
\usepackage[usenames, dvipsnames]{color}
\usepackage{soul} 
\usepackage{pslatex}
\usepackage{tabu}
\usepackage{makecell}
\usepackage{textcomp}
\usepackage{siunitx}
\usepackage{placeins}
\usepackage{xr}
\usepackage{amsmath}
\usepackage{natbib}

\title{Two-step solid-state synthesis of ternary nitride materials}

\author{Paul K. Todd}
\affiliation{National Renewable Energy Laboratory, Material Science Center, Golden, Colorado 80401, United States.}
\email{paul.todd@nrel.gov}
\author{M. Jewels Fallon}
\affiliation{Department of Chemistry, Colorado State University, Fort Collins, Colorado 80523-1872, United States}
\author{James R. Neilson}
\email{james.neilson@colostate.edu}
\affiliation{Department of Chemistry, Colorado State University, Fort Collins, Colorado 80523-1872, United States}
\author{Andriy Zakutayev}
\affiliation{National Renewable Energy Laboratory, Material Science Center, Golden, Colorado 80401, United States.}
\email{Andriy.Zakutayev@nrel.gov}

\begin{document}

\begin{tocentry}
	\begin{center}
			\includegraphics[height=3.5cm]{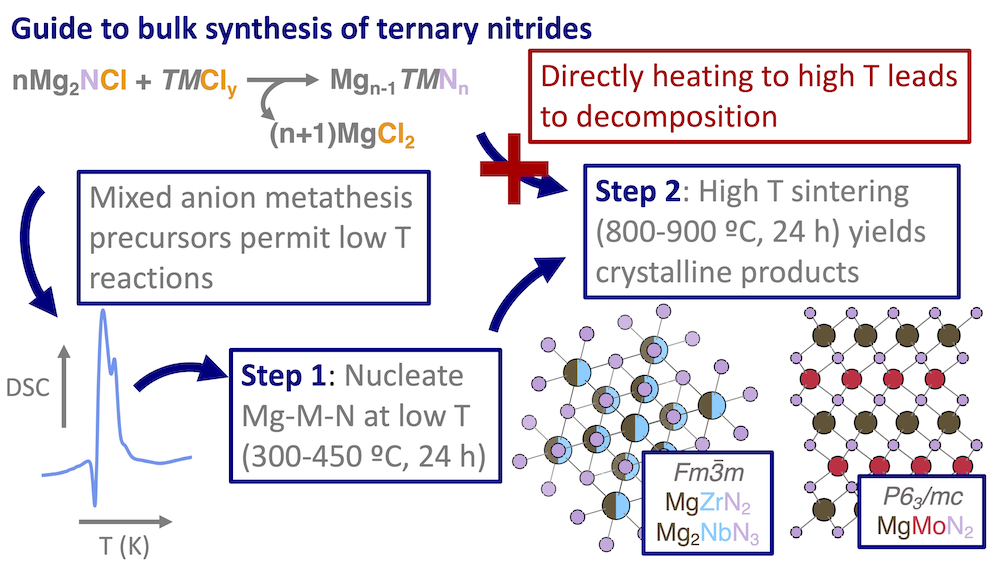}
		\end{center}
\end{tocentry}

\begin{abstract}
	Ternary nitride materials hold promise for many optical, electronic, and refractory applications yet their preparation via solid-state synthesis remains challenging. Often, high pressures or reactive gasses are used to manipulate the effective chemical potential of nitrogen, yet these strategies require specialized equipment. Here we report on a simple two-step synthesis using ion-exchange reactions that yield rocksalt-derived \ce{MgZrN2} and \ce{Mg2NbN3}, as well as  layered \ce{MgMoN2}. All three compounds show nearly temperature-independent and weak paramagnetic responses to an applied magnetic field at cryogenic temperatures indicating phase pure products. The key to synthesizing these ternary materials is an initial low-temperature step (300-450 \textcelsius{}) to promote \ce{Mg-\textit{M}-N} bond formation. Then the products are annealed (800-900 \textcelsius{}) to increase crystalline domains of the ternary product. Calorimetry experiments reveal that initial reaction temperatures are determined by phase transitions of reaction precursors, whereas heating directly to high temperatures results in decomposition. These two-step reactions provide a rational guide to material discovery of other bulk ternary nitrides.
\end{abstract}

Ternary metal nitrides remain under-explored as new functional inorganic materials,\cite{Zakutayev2016} even though a large number of new nitride compositions and structure types have been recently predicted.\cite{Hinuma2016, Gagne2020, Sun2019, Zakutayev2014a} The deficit in realized nitride products as compared to predicted materials stems from their difficult synthesis, with few successful reactions that yield nitride products. Generally, reactions must proceed at low temperatures, where dinitrogen (\ce{N2}) formation is less thermodynamically favorable,  or reactions must change the effective chemical potential within the reaction system through use of high pressures or reactive gasses like ammonia. Furthermore, a high number of potential binary metal nitride precursors are either refractory\cite{pierson2013handbook, braithwaite2013solid} or energetic\cite{Zeman2020, Odahara2018, Yin2015}, which further reduces the number of useful reactions. Therefore, identifying sources of reactive nitrogen that yield desired products under mild conditions is imperative for advancement in nitride material discovery. 

	\begin{figure}[ht]
		\begin{center}
			\includegraphics[width=7cm]{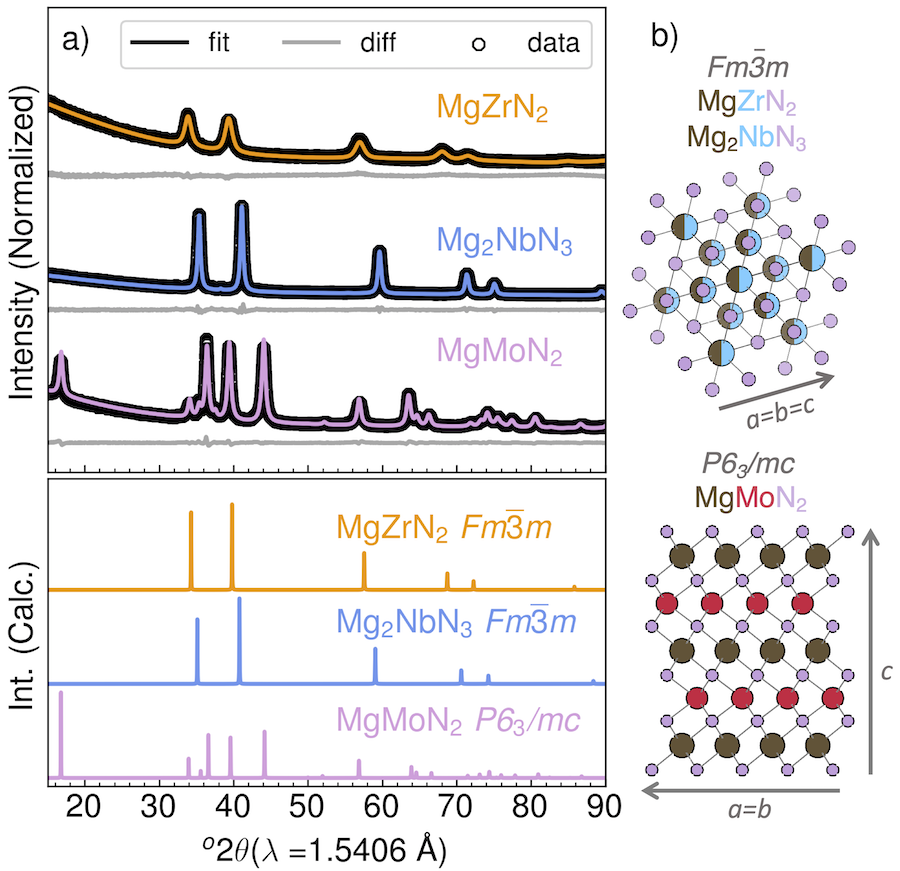}
			\caption{\label{FIG:PXRD} Crystallographic structure of three magnesium metal nitride products. a) Rietveld analysis of PXRD patterns of magnesium metal nitride products along with the simulated reference diffraction patterns and space group of each structure. b) Pictorial representations of the cubic rocksalt-derived structure for \ce{MgZrN2} and \ce{Mg2NbN3}, and a layered hexagonal crystal structure for \ce{MgMoN2}.} 
		\end{center}
	\end{figure} 
	
Various reactive nitride sources are used for ternary metal nitride synthesis. For thin film materials discovery and applications, an excited nitrogen plasma can be employed to deposit ternary metal nitrides ranging from rocksalt magnesium metal nitrides, such as \ce{MgZrN2} and \ce{Mg2NbN3}\cite{Bauers2019,Bauers2019a}, to wurtzite zinc metal nitrides such as \ce{Zn2SbN3} and \ce{Zn3MoN4}.\cite{Arca2018, Arca2019} In bulk form, ternary nitrides have been synthesized by high pressure metathesis\cite{Kawamura2021, KAWAMURA}, ammonolysis\cite{Elder1993}, ammonothermally\cite{Hausler2018}, and rarely from the elements under flowing \ce{N2}\cite{Verrelli2017}. For the ambient-pressure synthesis of magnesium metal nitrides, lower temperatures are required to avoid the loss of magnesium from the ternary products. For example, in the synthesis of layered \ce{MgMoN2}, high pressure autoclaves are used in conjunction with sodium azide at 700  \textcelsius{} to achieve crystalline products.\cite{Wang2012} As a gentler alternative, mixed anion magnesium chloride-nitride \ce{Mg2NCl} has been recently used for lowering reaction temperatures in the preparation of binary \ce{Mn3N2}\cite{Rognerud2019} and ternary \ce{Mg_{x}Zr_{1-x}N}, including \ce{MgZrN2}.\cite{Rom2021}
	
Here we describe the synthesis of three magnesium metal nitrides, where a transition metal halide (\ce{ZrCl4}, \ce{NbCl5}, \ce{MoCl5}) reacts with magnesium chloronitride\cite{Li2015a} to yield each magnesium metal nitride  products (\ce{MgZrN2}, \ce{Mg2NbN3}, \ce{MgMoN2}) and equivalent amounts of \ce{MgCl2} byproduct:
	
	\begin{align}
		\label{eq:MgMN}
		\ce{&\textit{n}Mg2NCl + MCl_{y} -> Mg_{n-1}MN_{n} + (\textit{n+1})MgCl2},
	\end{align}
	
\noindent These ternary metal nitrides are synthesized close to ambient pressure by two-step reactions, where precursors are first heated at relatively low-temperature (300-450 \textcelsius{}) to promote \ce{Mg-M-N} bond formation, and then the temperatures is raised (800-900 \textcelsius{}) to crystallize the ternary metal nitride products. Exothermic events occur near the first low-temperature step for each composition, indicating intermediate reactions that likely yield magnesium metal nitride products. As a result of these two-step reactions, \ce{MgZrN2} and \ce{Mg2NbN3} are observed in a rocksalt-derived structure while \ce{MgMoN2} adopts a layered hexagonal structure. Close to stoichiometric cation compositions of the ternary nitride products are consistent with their weak paramagnetic behavior, as opposed to a strong diamagnetic response characteristic of the binary nitride impurities. These results demonstrate a low-temperature two-step solid-state synthesis approach to ternary nitride materials

	\begin{table*}
	\begin{center}
		\caption{Refined structural parameters for the magnesium metal nitride products, measured chemical composition x=Mg/(Mg+M) in \ce{Mg_{x}M_{1-x}N_{y}} as compared to reference values and to literature data, and phase fraction of binary metal nitride impurity as measured from PXRD and VSM measurements.}
		\begin{tabular}{cccccccc}
			\hline \hline
			\multicolumn{4}{ c }{Structure (PXRD)} & \multicolumn{2}{ c }{\makecell{Composition \\ x=Mg/(Mg+M)}} & \multicolumn{2}{ c }{\makecell{Binary Nitride \\ Phase Fraction}} \\
			Compound \ & $a=b$ (\AA) & $c$ (\AA)  & $V$ (\AA$^{3}$) & \makecell{EDX$^{*}$ \\ } & PXRD & \makecell{PXRD\\mol~\% M$_x$N$_y$} & \makecell{VSM \\vol~\% SC M$_x$N$_y$} \\
			\hline
			\ce{MgZrN2}  & (\textit{Fm$\overline{3}$m}) & & & 0.50 & 0.50 & 0 & 0  \\ 
			Present  &  4.541(2) &  -- & 93.67(1) & 0.40 & 0.48(8) & -- & 2.2$\cdot10^-3$ \\
			Ref\cite{Bauers2019} & 4.54 &   -- & -- & 0.49 & -- & -- & --\\
			\hline
			\ce{Mg2NbN3} & (\textit{Fm$\overline{3}$m}) & & & 0.67 & 0.67 & 0 & 0  \\
			Present & 4.386(3) & -- & 84.39(2) & 0.64 & 0.60(6) & -- & 8.13$\cdot10^-3$\\
			Ref\cite{Bauers2019} & 4.37 & -- & -- & 0.68 & -- & --& -- \\
			\hline 
			\ce{MgMoN2} & (\textit{P6$_{3}$/mc}) & & & 0.50 & 0.50 & 0 & 0 \\
			Present & 2.924(1) & 10.4716(6) & 77.54(6) & 0.46 & 0.53(4) & 0.53(7) & 1.8$\cdot10^-3$\\
			Ref\cite{Verrelli2017} & 2.91059(3) & 10.5484(1) & -- & 0.52 & 0.48 & --& --\\
			\hline \\ 
		\end{tabular} \\
		$*$EDX errors range from 5-10 \%
		\label{table:table1}
	\end{center}
\end{table*}

For each ternary nitride synthesis, homogenously-mixed precursor powders were pelletized under argon and flame-sealed in evacuated quartz ampules. For reactions yielding \ce{MgZrN2} and \ce{Mg2NbN3}, ampules were heated in a muffle furnace to 450 \textcelsius{} for 24 h followed by a subsequent anneal at 800  \textcelsius{} for 24 h. Similarly, reactions yielding \ce{MgMoN2} were heated at 300 \textcelsius{} for 24 h then 900 \textcelsius{} for 24 h.  Cation compositions were measured using Energy Dispersive X-ray Spectroscopy (EDX). Powder X-ray diffraction (PXRD) was used to characterize each product's crystal structure and bulk magnetic susceptibility measurements using a vibrating-sample magnetometer (VSM) confirmed the product composition and purity. Temperature-dependent reactions profiles were determined from Differential Scanning Calorimetry (DSC) experiments. More detailed accounts of synthesis methods and characterization are provided in the Supplementary Information. 
	
Using these two-step metathesis reactions, three magnesium metal nitrides were selectively prepared and confirmed through diffraction. Figure~\ref{FIG:PXRD}a) depicts PXRD patterns of the reaction products, \ce{MgZrN2}, \ce{Mg2NbN3}, and \ce{MgMoN2}, after washing with anhydrous methanol to remove \ce{MgCl2} products. Quantitative crystallographic analysis using the Rietveld method reveals that the \ce{MgZrN2} and \ce{Mg2NbN3} crystallize in the rock-salt (\textit{Fm$\overline{3}$m}) structure as previously reported in thin film products\cite{Bauers2019}, whereas \ce{MgMoN2} forms in the layered hexagonal crystal structure (\textit{P6$_{3}$/mc})\cite{Verrelli2017}, as illustrated in Figure~\ref{FIG:PXRD}c). The simulated XRD patterns are shown in Figure~\ref{FIG:PXRD}b) for comparison, while structural parameters for each product are listed in Table~\ref{table:table1} as compared to literature.
	
For the two rocksalt structures, \ce{MgZrN2} and \ce{Mg2NbN3}, the observed PXRD patterns support magnesium inclusion into the rocksalt structure by a change in the (111) peak intensity, which is indicative of less electron density of the  magnesium cation. Rietveld analysis permits refinement of the site occupancies of the 4a Wykoff position in the rocksalt structure which accounts for the change in relative peak intensity in Figure~\ref{FIG:PXRD}a), with the x=Mg/(Mg+M) values reported in Table~\ref{table:table1}. For \ce{Mg_{x}Zr_{1-x}N_{y}} and \ce{Mg_{x}Nb_{1-x}N_{y}}, the cation concentrations fall within limits previously reported for these cation-disordered solid-solutions \cite{Bauers2019, Rom2021, Bauers2019a}. However, \ce{Mg_{x}Zr_{1-x}N_{y}} (x=0.48) and \ce{Mg_{x}Nb_{1-x}N_{y}} (x=0.60) reported here reveal a deficiency of magnesium as compared to ideal values (x=0.50 and x=0.67 respectively). Furthermore, the peak width for these rocksalt phases broadens with increasing cation site disorder, which has been previously observed for the solid-solution \ce{Mg_{x}Zr_{1-x}N}.\cite{Rom2021} 
	
Fitting the layered hexagonal structure of \ce{MgMoN2} using Rietveld refinement reveals an absence in intensity in the (0~0~\textit{l}) family of reflections relative to peaks associated with atoms in the (\textit{h}~0~1), as similarly observed for structurally analogous \ce{MnMoN2}.\cite{Bem1996} This observation can be explained by either disorder in the (0~0~1) direction of the \ce{MgMoN2} layers or preferred orientation of crystallites in the (1~0~1) direction. During the Rietveld analysis, applying preferred orientation in the (1~0~1) direction accounts for the increase in intensity of these reflections relative to the  (1~0~\textit{l}) family of peaks. Furthermore, there is a contraction of the c-axis (Table~\ref{table:table1}), which could indicate the presence of smaller \ce{Mo^{5+}} cations in the nominally \ce{Mo^{4+}} site, likely due  excess magnesium incorporation.  Free refinement of each cation site in the \textit{P6$_{3}$/mc} lattice (Table~\ref{table:table1}) supports greater magnesium content than molybdenum in \ce{Mg_{x}Mo_{1-x}N_{y}} (refined x=0.53 compared to x=0.50 reference value), along with some cation deficiency on the molybdenum site with (Mg+Mo)/(Mg+Mo+N)=0.85 compared to the 1.00 reference value. 
	
The relative cation composition in these \ce{Mg_{x}Zr_{1-x}N_{y}}, \ce{Mg_{x}Nb_{1-x}N_{y}} and \ce{Mg_{x}Mo_{1-x}N_{y}} materials, where x$=$Mg$/$(Mg$+$M), was confirmed by EDX analysis. For these metals, the EDX peak intensities are high enough to provide reasonable error, whereas nitrogen and oxygen differentiation is not as facile due to low signal to noise ratio in the low-energy part of the spectrum, as well as high background oxygen counts from the substrate. As presented in Table~\ref{table:table1}, the EDX results show magnesium and transition metal compositions that fall within the limits determined from our XRD refinement and previously reported in other publications\cite{Bauers2019, Rom2021, Bauers2019a, Verrelli2017}, although for the rocksalt products this ratio is sub-stoichiometric with regards to magnesium. Therefore, it is likely that some magnesium is lost during the reaction due to the thermal decomposition of \ce{Mg2NCl} at higher temperatures, despite intentional excess of this precursor in the reactions. In all reactions, a metal deposit is present on the quartz ampule, supporting the reduction of magnesium and formation of \ce{N2}. 
	
\begin{figure}[ht]
		\begin{center}
			\includegraphics[width=7cm]{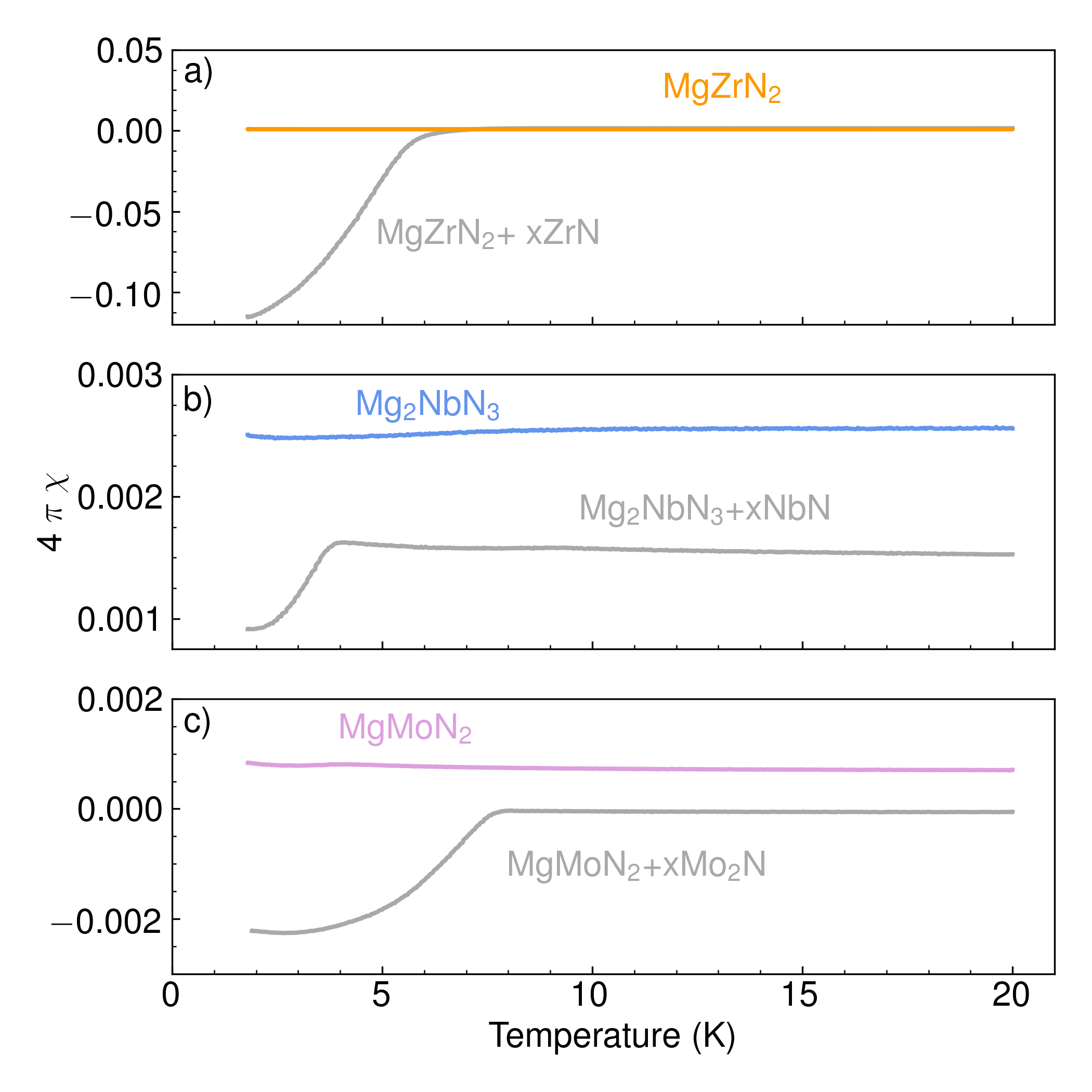} 
			\caption{\label{FIG:Mag} Temperature dependent magnetic susceptibility of magnesium metal nitride ternary products: a)\ce{MgZrN2}, b) \ce{Mg2NbN3}, c) \ce{MgMoN2}. The phase pure products washed with dry methanol are shown in contrast to samples with magnesium leached out by 1M nitric acid, leading to binary transition metal nitride impurities with clear superconducting transitions.DC magnetization data were collected using a measurement field of H $=$ 20 Oe.}
		\end{center}
\end{figure} 
	
To further evaluate the phase purity of our Mg-M-N products, magnetic susceptibility measurements were performed. The results in Figure~\ref{FIG:Mag} exhibit weak paramagnetic behavior ($\chi~>0$) that supports the compositions presented in Table~\ref{table:table1}. For each of these ternary Mg-M-N products, a binary metal nitride or oxynitride impurity (ZrN, NbN, \ce{Mo2N}) should produce a diamagnetic response from a superconducting transition, which is virtually absent in Figure~\ref{FIG:Mag} for the samples reported in Table~\ref{table:table1}. To illustrate the effect of even small fractions of binary nitride impurities, reaction products were treated with 1M nitric acid in attempts to leach out magnesium. These leaching experiment (Figure~\ref{FIG:Mag}) lead to a clear decrease in the magnetic susceptibility at low temperatures indicative of a superconducting transition in the binary impurity. It should be noted that the pure products washed with dry methanol (Fig.~\ref{FIG:Mag} and Table~\ref{table:table1}) do exhibit a very small superconducting transition (Fig.~\ref{SIFIG:Mag}) corresponding to <0.001 vol~\% impurity, yet these values are significantly lower in superconducting phase fraction than the products leached with nitric acid. 
	
The described synthesis conditions in Equation~\ref{eq:MgMN} require two-step temperature profiles where an initial temperature (T$_{rxn}$) promotes metal nitride bond formation, followed by a higher crystallization temperature (T$_{crst}$). Figure~\ref{SIFIG:Controls}~depicts PXRD patterns of the unwashed products observed when using these two-step heating profiles compared to directly heating to T$_{crst}$. When heated directly to 800~\textcelsius{}, the rocksalt \ce{MgZrN2} product observes a clear shift in lattice parameter (Fig.~\ref{SIFIG:Controls}a) towards ZrN paired with an increase in the relative intensity of the (1 1 1) peak, supporting a loss of Mg. For \ce{Mg2NbN3} products, heating directly to 800 \textcelsius{} results in broad peaks in the PXRD pattern (Fig.~\ref{SIFIG:Controls}b) with a shift towards a smaller lattice parameters than the reaction product via two-step heating schedule. The calculated ground state lattice parameter of 4.42 \AA~is larger than binary NbN, yet thin film \ce{Mg2NbN3} reports a lattice parameter of 4.37 \AA.\cite{Bauers2019} For reactions yielding \ce{MgMoN2}, directly heating above 800 \textcelsius{} yields more \ce{Mo2N} than \ce{MgMoN2}, whereas the described two-step heating profile increases the yield of \ce{MgMoN2} as seen in Figure~\ref{SIFIG:Controls}c. 
	
\begin{figure}[ht]
	\begin{center}
		\includegraphics[width=7cm]{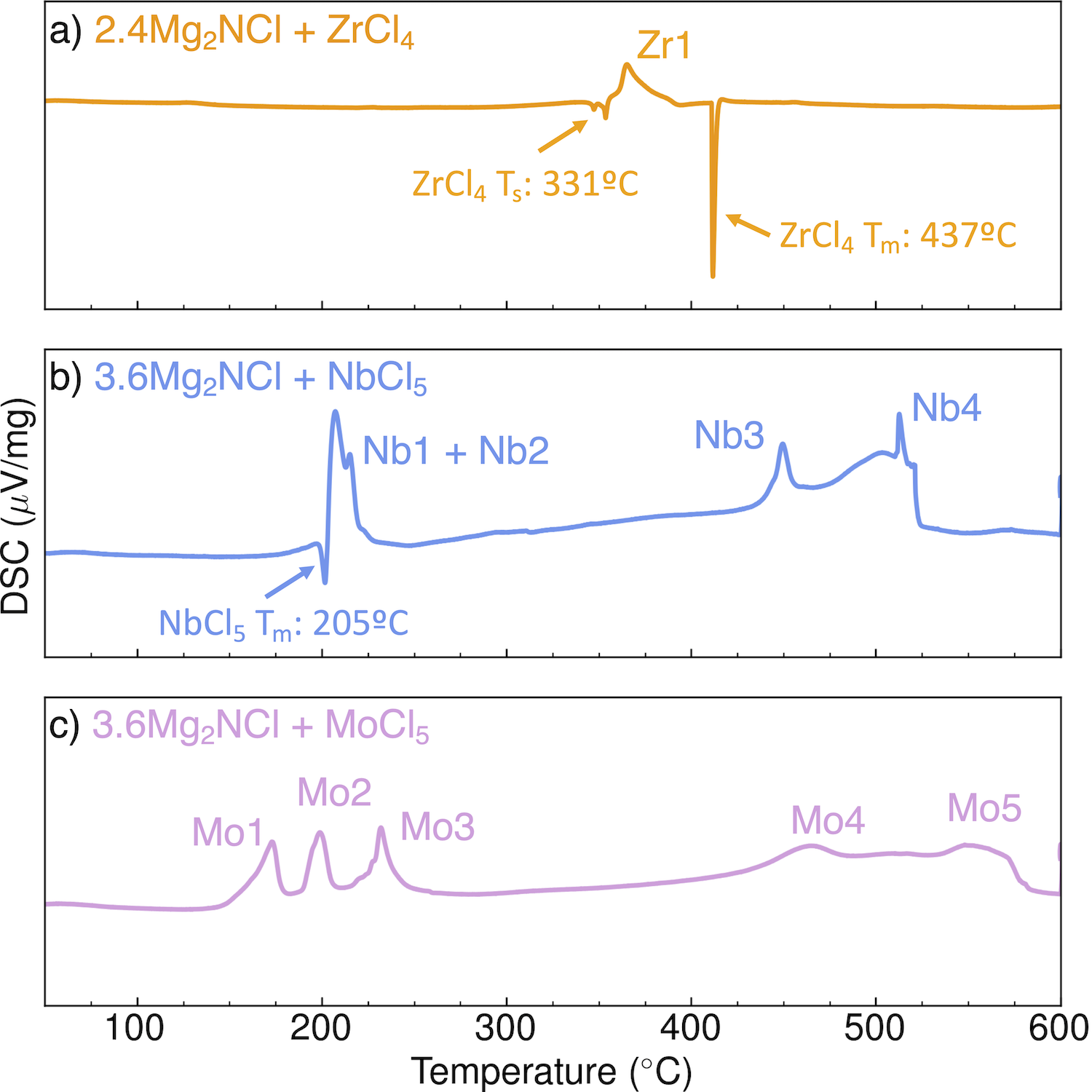}
		\caption{\label{FIG:DSC} Synthesis reaction pathway from DSC measurements for a) \ce{MgZrN2}, b) \ce{Mg2NbN3}, and c) \ce{MgMoN2}. The endotherms correlate with transitions of metal halide precursors and the exotherms  correspond to bond formation of ternary products.}
	\end{center}
\end{figure}

To gain  insight into the low-temperature reaction pathway, we performed DSC experiments presented in Figure~\ref{FIG:DSC}. These DSC results reveal new low-temperature exothermic reactions paired with known endothermic phase transitions of the respective transition metal halide precursors. \ce{Mg2NCl} does not have a phase transition below 600 \textcelsius{} suggesting that observed exotherms are attributed to the formation of \ce{MgCl2}, \ce{Mg-M-N} products, or unknown intermediate species. For the reactions yielding \ce{MgZrN2} (Fig.~\ref{FIG:DSC}a), there is an exotherm observed after the sublimation temperature of \ce{ZrCl4} at 331 \textcelsius{} (Zr1: 366 \textcelsius{}). At 411 \textcelsius{} a large endothermic inflection is observed, which we attribute to the pressure-induced melting of \ce{ZrCl4} from the gaseous state near 437 \textcelsius{}.\cite{Tetrachlorides1953a} For reactions yielding \ce{Mg2NbN3} (Fig.~\ref{FIG:DSC}b), a similar exothermic peak is observed after the melting point of \ce{NbCl5} at 205 \textcelsius{} (Nb$_{1}$:208 \textcelsius{}; Nb2: 216 \textcelsius{}), with two additional exotherms, Nb3: 450 \textcelsius{} and Nb4: 513 \textcelsius{}, also observed. For the \ce{MgMoN2} reaction in Figure~\ref{FIG:DSC}c), no phase transition endotherm is observed for \ce{MoCl5} at the expected melting point of 194 \textcelsius{}, yet a triplet of exothermic peaks are observed near this transition temperature (Mo1-3: 174, 200, and 233 \textcelsius{}). Furthermore, there are two additional broad exotherms at higher temperatures (Mo4: 465 \textcelsius{} and Mo5: 550 \textcelsius{}).
	
Using the measured temperatures of relevant exothermic peaks in Figure~\ref{FIG:DSC}, control reactions were carried out targeting three T$_{rxn}$: 300, 450, 600 \textcelsius{}, and two T$_{crst}$: 800, 900 \textcelsius{} to evaluate the effect of temperature on the reaction products. Figure~\ref{SIFIG:MgMoN2}a~depict the changes in hexagonal lattice parameters as a function of heating schedule for \ce{MgMoN2} products. Here the proposed T$_{rxn}$ of 300 \textcelsius{} yields lattice parameters most similar to \ce{MgMoN2}.\cite{Verrelli2017} As T$_{rxn}$ increases, a-axis lattice parameter remains constant, whereas c-axis lattice parameter decreases. Additionally, Figure~\ref{SIFIG:MgMoN2}a~further supports increased \ce{MgMoN2} yields at a lower T$_{rxn}$ whereas higher initial temperatures result in greater \ce{Mo2N} yields. Contrary to the lower T$_{crst}$ of the rocksalt yielding reactions, \ce{MgMoN2} product yields increase at 900 \textcelsius{}, albeit only with a two-step temperature profile.  

For rocksalt \ce{MgZrN2} and \ce{Mg2NbN3} products from these control reactions, Rietveld analysis was used to quantify changes in lattice parameter (Fig.~\ref{SIFIG:MgMoN2}a), as well as the changes in peak intensity relative to the (1 1 1) by allowing the magnesium to transition metal ratio to openly refine (Fig.~\ref{SIFIG:OccupancyVSHeating}). For \ce{MgZrN2}, proposed heating schedule results in the smallest lattice parameter and largest Mg concentration in Figure~\ref{SIFIG:Rocksalt}a. Additionally, the lower T$_{crst}$ results in peak broadening as calculated in Figure~\ref{SIFIG:Rocksalt}b, which also supports increased cation disorder in these rocksalt structures. For \ce{Mg2NbN3} the change in lattice parameter in Figure~\ref{SIFIG:Rocksalt}a~is less indicative of increased Mg content, yet changes in peak shape in Figure ~\ref{SIFIG:Rocksalt}b) support a similar trend of increased magnesium content with broadened peak shape, which requires a lower T$_{crst}$. 

The collective results reveal a synthesis approach to ternary magnesium metal nitrides at low temperatures and ambient pressures (Figure~\ref{FIG:PXRD}). A key to this two-step process is the dependence on a low-temperature reaction  T$_{rxn}$. We suggest that T$_{rxn}$ yields \ce{Mg-\textit{M}-N} bond formation as evidenced by the numerous exothermic events from DSC (Fig.~\ref{FIG:DSC}), and the absence of ternary products from direct heating (Fig.~\ref{SIFIG:Controls}). This low-temperature reaction step at T$_{rxn}$ ensures that solid-state diffusion can proceed below the temperature where product decomposition is observed, due to the overall small changes in formation energies and the increasing entropic driving force for \ce{N2} formation.\cite{Sun2017a, Sun2019}. This low-temperature reaction pathway is facilitated by the low melting points of the transition metal halide precursors, according to the DSC measurements in Figure~\ref{FIG:DSC}. These transition metal halides form monomeric or dimeric species as they melt\cite{BEVERIDGE1967, FAIRBROTHER1967, FERGUSSON1967a}, which reduces diffusion lengths at the reaction interface, thus ensuring that necessary ion-exchange occurs to yield \ce{Mg-M-N} intermediate phases or poorly crystalline products. Heating to higher temperatures T$_{crst}$ too quickly results in deleterious sublimation and decomposition of these halide precursors. Thus, a higher temperature T$_{crst}$ may be required to increase the crystallinity of the products, yet the T$_{rxn}$ reaction temperature is the most likely ``rate-limiting" step in this two-step reaction pathway.
	
The presented reaction conditions are benign and can be  performed in a traditional solid-state chemistry laboratory, thus increasing their utility in targeting other metal nitride compositions by multiple research groups. Previous studies on the synthesis of magnesium metal nitrides have employed custom high-pressure reactors,\cite{KAWAMURA, Kawamura2021} or specialized deposition chambers \cite{Bauers2019,Bauers2019a}. By starting with the mixed anion \ce{Mg2NCl} as a precursor, the reaction pathway does not proceed via a rapid propagation, as observed when starting with more energetic precursors such as alkali azides or alkali earth nitrides\cite{Zeman2020, Parkin2013} and diffusion-limited products and binary metal nitrides observed in some metathesis reactions\cite{Parkin2013, Gillan1996} are avoided. Furthermore, the presented reactions avoid toxic environments, like ammonia or amide-based mineralizers\cite{Hausler2018, Richter2014, Hausler2017}, that decompose under elevated temperatures and require careful safety considerations and custom equipment. Similar to the abundance of metal halide precursors, numerous metal chloronitride phases exist and are easily synthesized.\cite{Strahle, Hadenfeldt1987, Marx1997} For example, using \ce{Zn2NX} (X=Cl, Br, I) precursors to synthesize zinc metal nitrides such as \ce{Zn2NbN3} and  \ce{ZnZrN2}, which exhibit  high-temperature loss of zinc from the reaction products\cite{Liu2013, Zhao2021, Zakutayev2021,Woods-Robinson2020} is a logical next step in the discovery of new ternary metal nitride materials. 
	
	
In summary, we report on the bulk solid-state synthesis of three magnesium metal nitrides -- \ce{MgMoN2} with layered hexagonal structure, and \ce{MgZrN2}, \ce{Mg2NbN3} with rocksalt-derived structure -- using two-step low-temperature ion-exchange reactions. An initial low-temperature reaction of the precursors yields the magnesium metal nitride product, and is followed by a high temperature step to increase the product crystallinity measured by PXRD, with cation stoichiometry confirmed by EDX, and phae purity supported by magnetic susceptibility measurements. Characterizing this reaction pathway using DSC reveals multi-step crystallization that occur at low temperatures, which we attribute to the formation of the intermediate ternary product with short-range ternary metal nitride bonds but without long-range crystallographic order. In contrast, by heating too rapidly the precursors before they can successfully nucleate the magnesium metal nitride products results in a net loss of Mg and N at high temperature. The results presented here indicate that this low-temperature ambient-pressure approach can be used to synthesize other ternary nitride materials.
	
This work was performed at the National Renewable Energy Laboratory, operated by Alliance for Sustainable Energy, LLC, for the U.S. Department of Energy (DOE) under Contract No. DE-AC36-08GO28308. Funding provided by Office of Science (SC), Office of Basic Energy Sciences (BES), Materials Chemistry program, as a part of the Early Career Award “Kinetic Synthesis of Metastable Nitrides” (synthesis, composition, and structure measurements at NREL). Magnetic measurements at CSU were supported by the National Science Foundation (DMR-1653863). We would like to thank Bobby To and Max Schulze for assistance with SEM-EDX measurements. The authors acknowledge Annalise Maughan for her generous use of glovebox and furnace space. The views expressed in this article do not necessarily represent the views of the DOE or the U.S. Government.

\begin{suppinfo}
	The Supporting Information contains experimental methods and additional results from control reactions. 
\end{suppinfo}

\bibliographystyle{achemso}
\bibliography{library.bib}

\end{document}